\documentclass{llncs}

\usepackage[utf8]{inputenc}
\usepackage{amssymb,amsmath,stmaryrd}
\usepackage{array} 
\usepackage{blkarray} 
\usepackage{tikz} 
\usetikzlibrary{arrows,shapes,snakes,backgrounds}
\usepackage[colorlinks=true, citecolor=black,linkcolor=black]{hyperref}
\usepackage{algorithm}
\usepackage[noend]{algorithmic}
\usetikzlibrary{arrows,shapes,snakes,backgrounds}
\usepackage{paralist}
\makeatletter
\renewcommand*{\@fnsymbol}[1]{\ensuremath{\ifcase#1\or \star\or \dagger\or \ddagger\or
       \mathsection\or \mathparagraph\or \|\or **\or \dagger\dagger
       \or \ddagger\ddagger \else\@ctrerr\fi}}
\makeatother

\newcommand{\states}{\ensuremath{S} }
\newcommand{\statesOne}{\ensuremath{S_{1}} }
\newcommand{\statesTwo}{\ensuremath{S_{2}} }
\newcommand{\initState}{\ensuremath{s_{init}} }
\newcommand{\edges}{\ensuremath{E} }
\newcommand{\dimension}{\ensuremath{k} }
\newcommand{\weight}{\ensuremath{w} }

\newcommand{\game}{\ensuremath{G} }

\newcommand{\parity}{\ensuremath{p} }
\newcommand{\parityFull}{\ensuremath{p : \states \rightarrow \nat} }
\newcommand{\gameParFull}{\ensuremath{G = \left( \statesOne, \statesTwo, \initState, \edges, \dimension, \weight, \parity\right)} }

\newcommand{\gamePar}{\ensuremath{G} }
\newcommand{\integ}{\ensuremath{\mathbb{Z}} }
\newcommand{\nat}{\ensuremath{\mathbb{N}} }

\newcommand{\player}{\ensuremath{\mathcal{P}} }

\newcommand{\play}{\ensuremath{\pi} }
\newcommand{\plays}{\ensuremath{\textsf{Plays}(\game)} }

\newcommand{\prefixesArg}[1]{\ensuremath{\textsf{Prefs}_{#1}(\game)} }

\newcommand{\prefixesArgPar}[1]{\ensuremath{\textsf{Prefs}_{#1}(\gamePar)} }
\newcommand{\prefix}{\ensuremath{\rho} }

\newcommand{\strat}{\ensuremath{\lambda} }

\newcommand{\node}{\ensuremath{\varsigma} }

\floatstyle{plain}
\newfloat{myalgo}{tbhp}{mya}
{\begin{myalgo}[#1]
\centering
\begin{minipage}{#2}
\begin{algorithm}[H]}%
{\end{algorithm}
\end{minipage}
\end{myalgo}}

\title{Automated synthesis of reliable and efficient systems through game theory: a case study}
\author{Mickael Randour\inst{1}$^{,}$\thanks{Author supported by F.R.S.-FNRS. fellowship.} }

\institute{
Institut d'Informatique, Université de Mons (UMONS), Belgium
}

\begin{document}

\maketitle

\begin{abstract}
Reactive computer systems bear inherent complexity due to continuous interactions with their environment. While this environment often proves to be uncontrollable, we still want to ensure that critical computer systems will not fail, no matter what they face. Examples are legion: railway traffic, power plants, plane navigation systems, etc. Formal verification of a system may ensure that it satisfies a given specification, but only applies to an already existing model of a system. In this work, we address the problem of synthesis: starting from a specification of the desired behavior, we show how to build a suitable system controller that will enforce this specification. In particular, we discuss recent developments of that approach for systems that must ensure Boolean behaviors (e.g., reachability, liveness) along with quantitative requirements over their execution (e.g., never drop out of fuel, ensure a suitable mean response time). We notably illustrate a powerful, practically useable algorithm for the automated synthesis of provably safe reactive systems.
\end{abstract}

\section{Context}
Nowadays, more and more aspects of our society depend on \textit{critical reactive systems}, i.e., systems that continuously interact with their uncontrollable environment. Think about control programs of power plants, ABS for cars or airplane and railway traffic managing. Therefore, we are in dire need of systems capable of sustaining a safe behavior despite the nefarious effects of their environment.

Good developers know that testing do not capture the whole picture: never will it \textit{proves} that no bug or flaw is present in the considered system. So for critical systems, it is useful to apply \textbf{formal verification}. That means using \textit{mathematical tools} to prove that the system follows a given specification which models desired behaviors. While verification applies \textit{a posteriori}, checking that the formal model of a system satisfies the needed specification, it is most of the time desirable to start \textit{from} the specification and automatically build a system from it, in such a way that desired properties are proved to be maintained in the process. This \textit{a priori} process is known as \textbf{synthesis}.
 
The mathematical framework we use is \textbf{game theory}. It is a wide field with extensive formal bases and applications in numerous disciplines as diverse as economics, biology, operations research and, of course, computer science. Games model interactions between cooperating and/or competiting players who play to the best of their abilities in order to satisfy individual or common objectives. While interesting works of Borel \cite{borel1938applications} and even Cournot \cite{cournot1838recherches} precede them, von Neumann and Morgenstern are generally considered as the ``Founding Fathers of (Modern) Game Theory'' through their 1944 book titled \textit{Theory of Games and Economic Behavior} \cite{von1944theory}.

Roughly speaking, we consider a reactive system as a player (player 1), and his uncontrollable environment as its adversary (player 2). We model their interactions in a game on a graph, where vertices model states of the system and its environment, and edges model their possible actions. Constructing a correct system controller then means devising a \textit{strategy} (i.e., a succession of choices of actions) for player 1 such that, whatever the strategy of player 2, the outcome of the play satisfies the specification. Such game-theoretic formulations have proved useful for synthesis \cite{Church62,pnueli_POPL89,RamadgeWonham87}, verification~\cite{AHK02}, refinement~\cite{FairSimulation}, and compatibility 
checking \cite{InterfaceTheories} of reactive systems, as well as 
in analysis of emptiness of automata~\cite{Thomas97}.

\tikzstyle{decision} = [diamond, draw,
    text width=4.5em, text badly centered, node distance=3cm, inner sep=0pt]
\tikzstyle{block} = [rectangle, draw, 
    text width=5em, text centered, rounded corners, node distance=3.5cm, minimum height=4em]
\tikzstyle{none} = []
\tikzstyle{line} = [draw, -latex']
\tikzstyle{cloud} = [draw, ellipse, node distance=5cm, text badly centered, minimum height=2em,text width=6em]
    
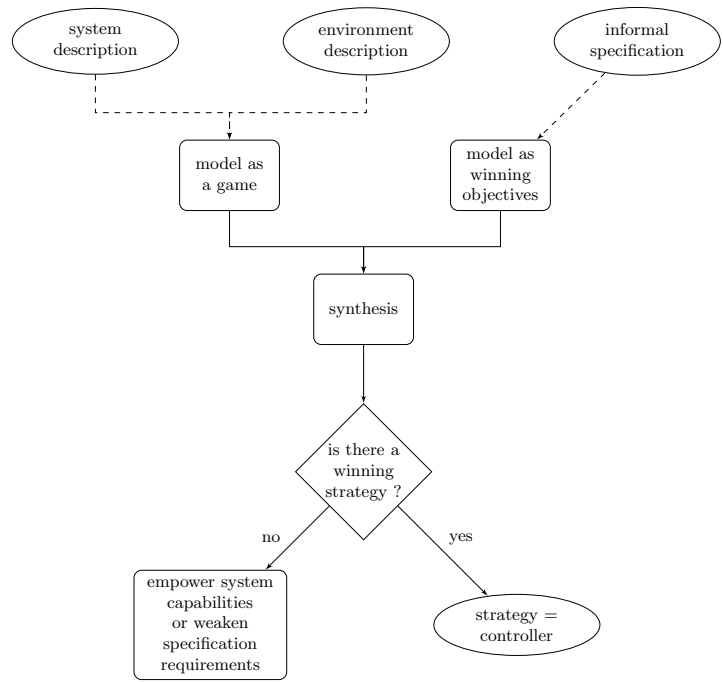
\begin{figure}[htb]
  \centering
  \resizebox{9.5cm}{!}{
\begin{tikzpicture}[node distance = 2cm, auto]
    \node [cloud] (sysdesc) {system description};
    \node [cloud, right of=sysdesc] (envdesc) {environment description};
    \node [cloud, right of=envdesc] (infspec) {informal specification};
    \node [block, below right of=sysdesc] (game) {model as a game};
    \node [block, right of=game, node distance=5cm] (objec) {model as winning objectives};
    \node [block, below right of=game] (synth) {synthesis};
    \node [decision, below of=synth] (iswin) {is there a winning strategy ?};
    \node [block, below left of=iswin,text width=8em, node distance=4cm] (revise) {empower system capabilities or weaken specification requirements};
    \node [cloud, below right of=iswin, node distance=4cm] (controller) {strategy = controller};
    \path [line,dashed] (sysdesc) |- ([yshift=0.5cm]game.north) -- (game);
    \path [line,dashed] (envdesc) |- ([yshift=0.5cm]game.north) -- (game);
    \path [line,dashed] (infspec) -- (objec);
    \path [line] (game) |- ([yshift=0.5cm]synth.north) -- (synth);
    \path [line] (objec) |- ([yshift=0.5cm]synth.north) -- (synth);
    \path [line] (synth) -- (iswin);
    \path [line] (iswin) -- node [left, xshift=-0.2cm] {no}(revise);
    \path [line] (iswin) -- node {yes}(controller);
\end{tikzpicture}}
      \caption{Controller synthesis through game theory: process.}
\label{fig:process}
  \end{figure}

In this paper, we do not address the full theoretical deepness of such an approach but rather try to motivate and illustrate its usefulness towards an audience who may not be familiar with it. To that end, we discuss a motivating toy example. First, we present the informal description of a reactive system and the behavior it should enforce. Second, we show how to use the game-theoretic framework to model its relationship with its environment and formalize the desired specification. Third, we use the sound theory of synthesis and exhibit a suitable controller that ensures satisfaction of the specification. Our discussion is mostly high level and intuitive.

A wide variety of games (and thus system models) have been studied recently, with diverse enforceable behaviors \cite{bloem_CAV09,bouyer_ATVA11,brazdil_ICALP10,chatterjee_ICALP10,chatterjee_FSTTCS10,chatterjee_LICS05,fahrenberg_ICTAC11,Mar98}. In this work, we will focus on systems that must satisfy \textit{qualitative behaviors} (e.g., always eventually granting requests, never reaching a deadlock) along with multiple \textit{quantitative requirements} (e.g., maintaining a bound on the mean response time, never running out of energy). In particular, we illustrate recent results of Chatterjee \textit{et al.} \cite{CRR12} that are the first to provide a synthesis algorithm for such games, as well as a deep study of the complexity of the synthesized controllers.

\section{Problem}
\label{sec:problem}
Consider the following running example. We want to synthesize a controller for a robotized lawnmower. This lawnmower is automatically operated, without any human intervention. We present its informal specification, as well as the effects the environment can have on its operation.

\begin{itemize}
\item In this partial, simplified specification, the gardener do not ask for the lawnmower to satisfy any bound on the frequency of grass-cuttings. However, as he wants that the grass does not grow boundlessly, the lawnmower should cut the grass infinitely often in the future (as if it stops someday, the grass will not stop growing from then on).
\item The lawnmower has an electric battery that can be recharged under sunshine thanks to solar panels, and a fuel tank that can only be filled when the lawnmower is back on its base. Both are considered unbounded to keep things simple.
\item The weather can be cloudy or sunny.
\item The lawnmower can refuel ($2$ fuel units) at its base under both weather conditions, but can only recharge its battery ($2$ battery units) when it is sunny. Resting at the base takes $20$ time units.
\item When cloudy, it can operate either under battery ($1$ battery unit) or using fuel ($2$ fuel units), both according to the same speed ($5$ time units). When sunny, the lawnmower may either cut the grass slowly, which always succeeds and consumes no energy (as the sun recharges the battery along the way), but takes $10$ time units. Or it may cut the grass fast, which consumes both $1$ unit of fuel and $1$ unit of battery, but only takes $2$ time units. 
\item When operating fast, the lawnmower makes considerably much noise, which may wake up the cat that resides in the garden and prompt it to attack the lawnmower. In that case, the grass-cutting is interrupted and the lawnmower goes back to its base, losing $40$ time units as repair is needed. The cat does not go out if the weather is bad.
\item As the gardener cannot benefit from his garden while the lawnmower is operating, he wants that the mean time required by actions of the lawnmower is less than $10$ time units.
\end{itemize}

While simple, this toy example already involves qualitative requirements (i.e., the grass should be mown infinitely often), along with quantitative ones. There are indeed three quantities that have to be taken into account: battery and fuel are energy quantities, which should never be exhausted, and time per action is a quantity which mean over an infinite operating of the lawnmower should be less than a given bound.

Given this informal description of the capabilities of the system and its environment, as well as the specification the system should enforce, we need to build a system controller that guarantees satisfaction of the specification.

\section{Modeling as a game}

\smallskip\noindent\textbf{Game.} We model the states and the interactions of the couple system/environment as a graph game where the system (here, the lawnmower) is player 1 and the environment is its adversary player 2. Formally, a \textit{game structure} is a tuple \gameParFull where (i) \statesOne and \statesTwo resp. denote the finite sets of \textit{states} belonging to player 1 and player 2, with $\statesOne \cap \statesTwo = \emptyset$; (ii) $\initState \in \states = \statesOne \cup \statesTwo$ is the initial state; (iii) $\edges \subseteq \states \times \states$ is the set of \textit{edges} s.t. for all $s \in \states$, there exists $s' \in \states$ s.t. $(s, s') \in \edges$; (iv) $k \in \nat$ is the \textit{dimension} of the weight vectors; (v) $\weight : \edges \rightarrow \integ^{\dimension}$ is the multi-weight labeling function; and (vi) $\parityFull$ is the priority function.

The game starts at an initial state, and if the current state is a 
player 1 (resp. player 2) state, then player 1 (resp. player 2) chooses an outgoing \textit{edge}. This choice is made according to a \textit{strategy} of the player: given the sequence of visited states, a strategy chooses an outgoing edge. For this case study, we only consider strategies that operate this choice deterministically. This process of choosing edges is repeated forever, and gives rise to an outcome of the game, called a {\em play}, that consists of the infinite sequence of states that are visited. Formally, a \textit{play} in \game is an infinite sequence of states $\play = s_{0}s_{1}s_{2}\ldots{}$ s.t. $s_{0} = \initState$ and for all $i \geq 0$, we have $(s_{i}, s_{i+1}) \in \edges$. The \textit{prefix} up to the $n$-th state of play $\play = s_{0}s_{1}\ldots{}s_{n}\ldots{}$ is the finite sequence $\play(n) = s_{0}s_{1}\ldots{}s_{n}$. Such a prefix $\play(n)$ belongs to player $i$, $i \in \lbrace 1, 2\rbrace$, if $s_{n} \in \states_{i}$. The set of plays of \game is denoted by $\plays$. The set of prefixes that belong to player $i$ is denoted by $\prefixesArg{i}$.
  
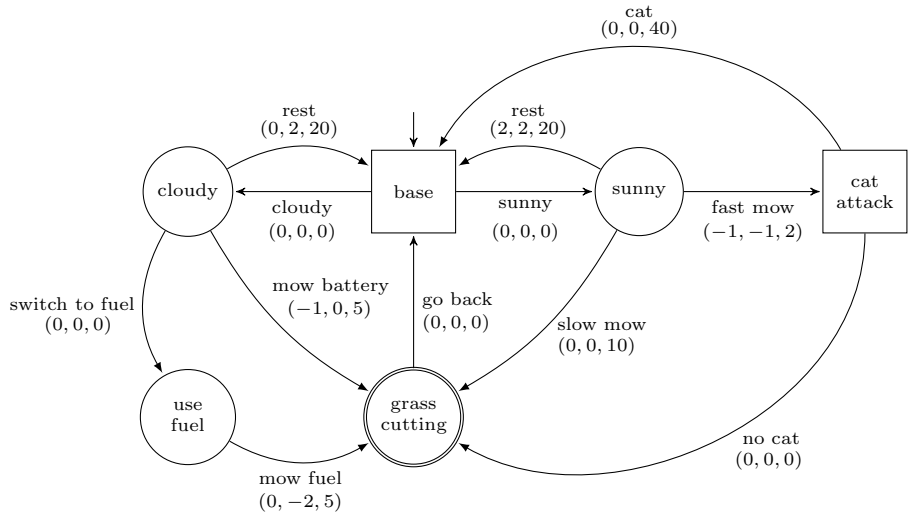
\begin{figure}[htb]
  \centering   
  \begin{tikzpicture}[->,>=stealth',shorten >=1pt,auto,node
    distance=2.5cm,bend angle=45,scale=0.75,font=\scriptsize]
    \tikzstyle{p1}=[draw,circle,text centered,minimum size=11mm, text width=9mm]
    \tikzstyle{p2}=[draw,rectangle,text centered,minimum size=11mm, text width=9mm]
    \node[p1]  (0)  at (0, 0) {cloudy};
    \node[p2]  (1) at (4, 0) {base};
    \node[p1]  (2)  at (8, 0) {sunny};
    \node[p2]  (3)  at (12, 0) {cat attack};
    \node[p1,double]  (4)  at (4, -4) {grass cutting};
    \node[p1]  (5)  at (0, -4) {use fuel};
    \coordinate[shift={(0mm,5mm)}] (init) at (1.north);
    \path
    (1) edge node [below] {cloudy} node [below, yshift=-0.3cm] {$(0,0,0)$} (0)
    (1) edge node [below] {sunny} node [below, yshift=-0.3cm] {$(0,0,0)$}  (2)
    (2) edge node [below] {fast mow} node [below, yshift=-0.3cm] {$(-1,-1,2)$} (3)
    (4) edge node [right] {go back} node [right, yshift=-0.3cm] {$(0,0,0)$} (1)
    (init) edge (1);
	\draw[->,>=latex] (0) to[out=300,in=150] node [right, xshift=-0.2cm, yshift=0.5cm] {mow battery} node [right, yshift=0.2cm] {$(-1,0,5)$} (4);
	\draw[->,>=latex] (0) to[out=240,in=120] node [left] {switch to fuel} node [left, yshift=-0.3cm, xshift=-0.3cm] {$(0,0,0)$} (5);
	\draw[->,>=latex] (5) to[out=330,in=210] node [below] {mow fuel} node [below, yshift=-0.3cm] {$(0,-2,5)$} (4);
	\draw[->,>=latex] (2) to[out=240,in=30] node [right] {slow mow} node [right,yshift=-0.3cm] {$(0,0,10)$} (4);
	\draw[->,>=latex] (3) to[out=270,in=330] node [right,xshift=0.2cm] {no cat} node [right,xshift=0.1cm,yshift=-0.3cm] {$(0,0,0)$} (4);
	\draw[->,>=latex] (0) to[out=30,in=150] node [above,yshift=0.3cm] {rest} node [above] {$(0,2,20)$} (1);
	\draw[->,>=latex] (2) to[out=150,in=30] node [above,yshift=0.3cm] {rest} node [above] {$(2,2,20)$} (1);
	\draw[->,>=latex] (3) to[out=120,in=60] node [above,yshift=0.3cm] {cat} node [above] {$(0,0,40)$} (1);
      \end{tikzpicture}
      \caption{Lawnmower game. Edges are fitted with tuples denoting changes in battery, fuel and time respectively.}
\label{fig:game}
  \end{figure}

Applying this formalism, we represent the lawnmower problem as the game depicted on Fig. \ref{fig:game}. Edges correspond to choices of the system or its environment and taking an edge implies a change on the three considered quantities, as denoted by the edge label. The \textit{grass-cutting} state is special as the specification requires that it should be visited infinitely often by a suitable controller.

\smallskip\noindent\textbf{Strategies.} Formally, a \textit{strategy} for $\player_{i}$, $i \in \lbrace 1, 2\rbrace$, in \gamePar is a function $\strat_{i} : \prefixesArgPar{i} \rightarrow \states$ s.t. for all $\prefix = s_{0}s_{1}\ldots{}s_{n} \in \prefixesArgPar{i}$, we have $(s_{n}, \strat_{i}(\prefix)) \in \edges$. The history of a play (i.e., the previously visited states and their order of appearance) may thus in general be used by a strategy to prescribe its choice. A strategy $\strat_{i}$ for $\player_{i}$ has \textit{finite memory} if the history it needs to remember can be bounded. In that case, the strategy can be encoded by a deterministic Moore machine. As discussed earlier, a strategy of player 1 (the lawnmower) provides a complete description of a controller for the system, prescribing the actions to take in response to any situation. Therefore, our task is to build a strategy that satisfies the specification.

\smallskip\noindent\textbf{Objectives.} To devise such a strategy, it is needed to formalize the specification as objectives of the game. The conjunction of objectives yields a set of winning plays that endorse the specification. A strategy of player 1 is thus said to be \textit{winning} if, \textit{against every possible strategy of the adversary}, the play induced by following this strategy belongs to the winning set of plays.

The informal specification developed in Section \ref{sec:problem} is encoded as the following objectives. We ommit technical details for the sake of this case study.

\begin{itemize}
\item \textit{Battery and fuel.} Both constitute energy types which quantities are never allowed to drop below zero. A play is thus winning for the \textit{energy objective} if the running sum of the weights encountered along it (i.e., changes induced by the taken edges) never drops below zero on any of the first two dimensions.
\item \textit{Mean action time.} The specification asks that the lawnmower spends no more than $10$ time units per action on average in the long run. That is, it is allowed to take more than $10$ time units on some actions, but the long-run mean should be below this threshold. Therefore, the \textit{mean-payoff objective} requires that the limit of the mean of the third-dimension weights over the prefix of a play is lower than $10$.
\item \textit{Infinitely frequent grass-cutting.} To satisfy this part of the specification, a strategy of player 1 must ensure that the grass-cutting state is visited infinitely often along the induced play. This is encoded as a \textit{Büchi objective} (or as a \textit{parity objective} via the priority function in the most general case).
\end{itemize}

\section{Synthesis}

\smallskip\noindent\textbf{Process.} Since our desire is to build practical real-world controllers, we are only interested in strategies that require \textit{finite} memory. From a theoretical standpoint, there exist classes of games where infinite memory may help to achieve better results (see for example \cite{chatterjee_FSTTCS10}), but infinite-memory strategies are of no practical use, as implementing a controller with infinite memory capabilities is obviously ruled out.

The core of the synthesis process depicted on Fig. 1 is thus to construct, if possible, a finite-memory strategy that ensures satisfaction of the previously defined objectives, as well as a corresponding initial value of the energy parameters, commonly referred to as \textit{initial credit}. That is because for the energy objectives, it is allowed to start the game with some finite quantity in stock, before taking any action. Think about starting a race with some fuel in your tank.

While of importance for the analysis of systems with both qualitative and quantitative requirements, the synthesis problem for the class of games that is used to model the lawnmower problem, i.e., games with parity and multi energy or mean-payoff objectives, has only been considered recently \cite{CRR12}. In this paper, the complexity of synthesized controllers is studied and it is shown that for some systems, exponentially complex controllers are needed to enforce the specification. Moreover, exponential size controllers are always sufficient, i.e., if no exponential controller is able to enforce the specification, then implementing more complex controllers is no help.

\smallskip\noindent\textbf{Result 1 (Induced by \cite[Theorem 1]{CRR12}).} \textit{Enforcing a specification combining both qualitative and quantitative aspects may require exponential size controllers in terms of memory requirements in the worst case.}

\smallskip
Interestingly, answering the question ``does there exist a finite-memory controller that satisfies a given specification ?'' was shown to be coNP-complete in \cite{chatterjee_FSTTCS10}. However, no deterministic algorithm was known to synthesize such a controller. Only quite recently, a practically implementable algorithm of \textit{optimal complexity} for the synthesis of specification-wise suitable controllers was presented in \cite{CRR12}. This algorithm is both symbolic and incremental, and uses compact representations of data sets, thus being an ideal choice for implementation into synthesis tools.

\smallskip\noindent\textbf{Result 2 (Induced by \cite[Theorem 2]{CRR12}).} \textit{The synthesis of controllers for systems with qualitative and quantitative requirements, such as the lawnmower, is in EXPTIME.}

\smallskip
This algorithm automatically builds suitable controllers with regard to the desired specification, if one is constructible. Therefore, it is the key tool in the synthesis process depicted on Fig. 1, and gives rise to an innovative and sound approach to the conception of provably safe reactive systems.

\bigskip\noindent\textbf{Lawnmower controller.} To conclude our case study, we exhibit a synthesized controller that enforces the desired specification. Notice that there may exist other acceptable controllers. The one we present here is quite simple but already asks for some memory (in the form of bookkeeping of battery and fuel levels). The controller implements the following strategy:
\begin{itemize}
\item Start the game with empty battery and fuel levels.
\item If the weather is sunny, mow slowly.
\item If the weather is cloudy,
\begin{itemize}
\item if there is at least one unit of battery, mow on battery,
\item otherwise, if there is at least two units of fuel, mow on fuel,
\item otherwise, rest at the base.
\end{itemize}
\end{itemize}

Notice that this strategy guarantees never running out of energy (which satisfies the energy objectives), induces infinitely frequent grass-cuttings (which satisfies the Büchi objective), and produces a play on which the mean time per action is less than $10$ against any strategy of the adversary (which satisfies the mean-payoff objective). In this sample controller, the lawnmower never uses the ``fast mow'' action as the adversary could very well play ``cat'' and prevent visit of the grass-cutting state.

\section{Conclusion}
Through this case study, we have discussed how the game-theoretic framework can help in the synthesis of controllers. We have intuitively introduced some of the key underlying concepts such as games, strategies, qualitative and quantitative objectives. We have also discussed the recent development of a practically useable algorithm for the automated synthesis of valid controllers \cite{CRR12}.

It is worthwhile noticing that automated synthesis suites for fragments of the presented formalism or similar logics are already in practical use, such as the LTL synthesis tool \texttt{Acacia+} for example \cite{BBFJR12} (which only applies to qualitative requirements). Thanks to the recent developments on the conjunction of qualitative and quantitative objectives \cite{CRR12}, such tools could very well be extended to encompass all the needed complexity for the specification of real-world systems. Such an approach should be a leading trend for the analysis and synthesis of provably safe controllers for reactive systems in the near future. This discussion illustrates its interest, while abstracting the sound theory underneath.

\bibliographystyle{plain}
\bibliography{bib}

\end{document}